\documentclass{article}

\usepackage{ijcai24}

\usepackage{times}
\usepackage{soul}
\usepackage{url}
\usepackage[hidelinks]{hyperref}
\usepackage[utf8]{inputenc}
\usepackage[small]{caption}
\usepackage{graphicx}
\usepackage{amsmath}
\usepackage{amsthm}
\usepackage{booktabs}
\usepackage{algorithm}
\usepackage[switch]{lineno}

\usepackage[table]{xcolor}
\usepackage{bm}

\usepackage{bbold}
\usepackage{multirow}
\usepackage{booktabs} 
\usepackage{subcaption}
\usepackage{graphicx}
\usepackage{float}

\usepackage{algorithm}
\usepackage{algpseudocode}

\usepackage{algpseudocode}
\newcommand\Algphase[1]{ \vspace*{-.5\baselineskip}\Statex\hspace*{\dimexpr-\algorithmicindent-2pt\relax}\rule{\columnwidth}{0.4pt}%
\Statex\hspace*{-\algorithmicindent}\textbf{#1}%
\vspace*{-.5\baselineskip}\Statex\hspace*{\dimexpr-\algorithmicindent-1pt\relax}\rule{\columnwidth}{0.4pt}}%

\linenumbers

\title{OUCopula: Bi-Channel Multi-Label Copula-Enhanced Adapter-Based CNN for Myopia Screening Based on OU-UWF Images}

\author{
Yang Li$^{2}$\footnote{Co-first authors.}
\and
Qiuyi Huang$^{1*}$\and
Chong Zhong$^1$\and
Danjuan Yang$^4$\and
Meiyan Li$^4$\and
A.H. Welsh$^3$\and
Aiyi Liu$^5$\and
Bo Fu$^{2\dagger}$\and
Catherine C. Liu$^{1}$\thanks{Co-corresponding authors.}\And
Xingtao Zhou$^{4\dagger}$ 
\affiliations
$^1$Department of Applied Mathematics, The Hong Kong Polytechnic University\\
$^2$School of Data Science, Fudan University\\
$^3$College of Business and Economics, Australian National University\\
$^4$Eye Institute and Department of Ophthalmology, Eye \& ENT Hospital, Fudan University\\
$^5$Biostatistis and Bioinformatics Branch, Eunice Kennedy Shriver National lnstitute of Child Health and Human Developnent
\emails
\{18110980006, fu\}@fudan.edu.cn,
charley.huang@connect.polyu.hk,
\{chzhong, macliu\}@polyu.edu.hk,
\{docdanjuanyang, limeiyan0406073, doctzhouxingtao\}@126.com,
Alan.Welsh@anu.edu.au,
liua@mail.nih.gov
}

\begin{document}
\nolinenumbers
\maketitle

\begin{abstract}

Myopia screening using cutting-edge ultra-widefield (UWF) fundus imaging is potentially significant for ophthalmic outcomes. Current multidisciplinary research between ophthalmology and deep learning (DL) concentrates primarily on disease classification and diagnosis using single-eye images, largely ignoring joint modeling and prediction for Oculus Uterque (OU, both eyes).
Inspired by the complex relationships between OU and the high correlation between the (continuous) outcome labels (Spherical Equivalent and Axial Length), we propose a framework of copula-enhanced adapter convolutional neural network (CNN) learning with OU UWF fundus images (OUCopula) for joint prediction of multiple clinical scores. 
We design a novel bi-channel multi-label CNN which can (1) take bi-channel image inputs subject to both high correlation and heterogeneity (by sharing the same backbone network and employing adapters to parameterize the channel-wise discrepancy), and (2) incorporate correlation information between continuous output labels (using a copula).
Solid experiments show that OUCopula achieves satisfactory performance in myopia score prediction compared to backbone models.  Moreover, OUCopula can far exceed the performance of models constructed for single-eye inputs. 
Importantly, our study also hints at the potential extension of the bi-channel model to a multi-channel paradigm and the generalizability of OUCopula across various backbone CNNs.

\end{abstract}

\section{Introduction}

Global ocular pathologies, a major cause of visual impairment, impact at least 2.2 billion people, with myopia affecting around 130 million individuals globally \cite{holden2016global}. 
Notably, high myopia is prevalent in Asian countries like China, affecting about 600 million people, including over 100 million students \cite{wang2022automatic}. 
Since high myopia can lead to complications like cataracts and retinal issues, improvements to myopia screening techniques can have a substantial impact on public health.
To aid in screening and detecting ocular diseases, different fundus imaging techniques have been developed. There are two primary types: conventional fundus images and ultra-widefield (UWF) fundus images. 
Conventional fundus images exhibit certain constraints including a restricted visual range of a mere 30$^{\circ}$--75$^{\circ}$ and the requirement of substantial photographic expertise and patient cooperation \cite{aggarwal2017ultra}. 
In contrast, UWF imaging provides an expansive 200$^{\circ}$ field of view, and does not require highly specialized ophthalmologists, making it suitable for telemedicine applications in regions lacking medical resources \cite{ohsugi2017accuracy}. 
Although currently UWF imaging is widely used in developed countries \cite{nagiel2016ultra}, the limited availability of retinal specialists and adequately trained UWF interpreters poses significant challenges to the implementation of UWF imaging within China and other developing countries. Consequently, there is a pronounced lack of high-quality annotated data and intelligent screening models for both medical applications and scientific research in these regions.

Recently, deep learning (DL) has made important contributions to the field of medical imaging for detecting myopia-related diseases and other retinal diseases \cite{devda2019pathological,rauf2021automatic}. Especially, different kinds of convolutional neural networks (CNN) have been developed, including ResNet, Vision Transformer, U-net, and Mask-RCNN, which have shown promising performance in various aspects of disease prediction, diagnosis, and segmentation \cite{yan2019learning,li2021deep,mohan2022vit}.

However, there is a paucity of research that considers the complex relationship between Oculus Uterque (OU, both eyes). The majority of existing studies treat the two images obtained from the left (OS) and right eye (OD) of the same patient as distinct entities, neglecting the inherent relationship between them \cite{he2021multi}. One study has shown a significant correlation in the progression of ophthalmic diseases between bilateral eyes \cite{ferris2005simplified}. This implies that integrating information from bilateral eyes for myopia prediction would be a more effective approach. Hence, in this work, we will model OU simultaneously, taking images of both eyes as inputs to the network, thereby transforming the model into a bi-channel CNN. 

\begin{figure}[tb]
    \centering
    \includegraphics[width=1.0\columnwidth]{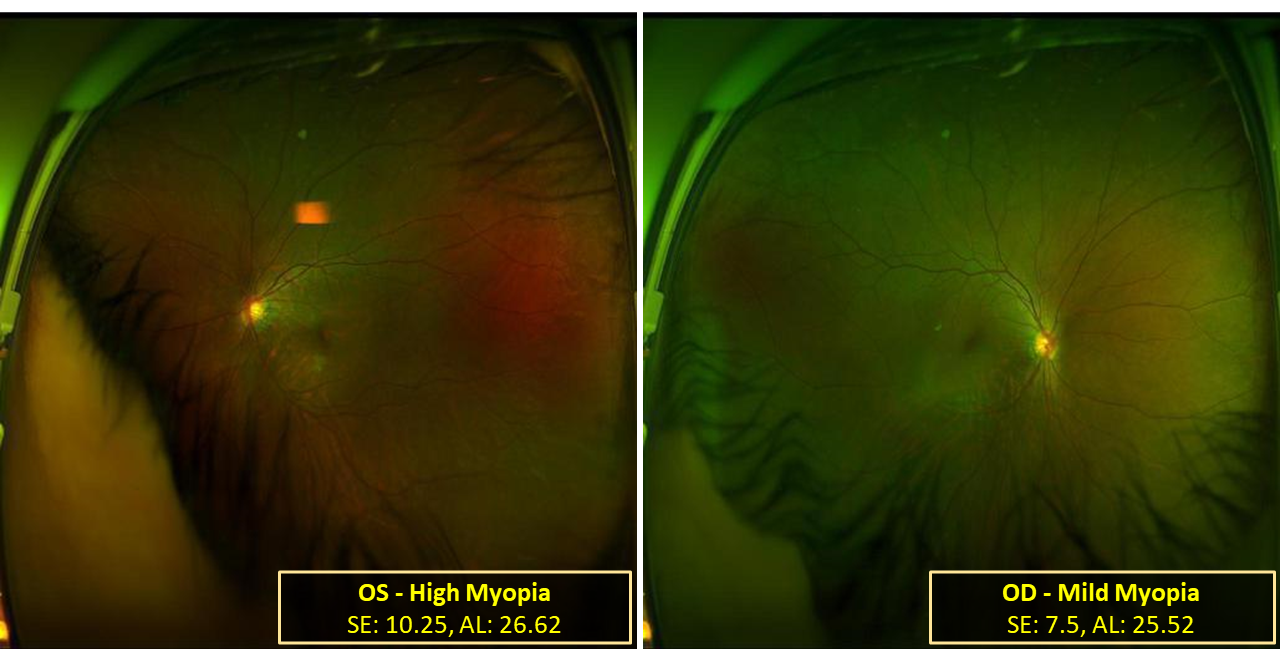}
    \caption{UWF fundus images of one patient indicating interocular asymmetries in myopia status}
    \label{fig:UWF_left_vs_right}
\end{figure}

However, heterogeneity also exists between OU; the so-called “interocular asymmetries” refer to asymmetrical or unilateral features between two eyes \cite{lu2022interocular}. In our study, the interocular asymmetry includes the scenario where the status of myopia in the two eyes of myopic patients is different, and it may even be that one eye is myopic while the other eye is not myopic. 
As Fig.~\ref{fig:UWF_left_vs_right} shows, the two eyes of the same patient can have very different myopia statuses: OS of this person is highly myopic, while OD is only mildly myopic. 
We have also done a Student's t-test, which shows that the difference in myopia status between OS and OD is statistically significant.
Researchers suggest that consideration of interocular asymmetries could reduce statistical bias and offer additional information about retinal diseases \cite{sankaridurg2013correlation,henriquez2015intereye}. 
This presents a challenge in the modeling process because both the strong correlation and the interocular asymmetry should be considered at the same time. 
Specifically, most parameters in the network should be shared to preserve the common features between two eyes, while the unique heterogeneous information contained in each eye also needs to be retained by distinct parts of the network. 
Inspired by the residual adapter introduced in \cite{rebuffi2018efficient} from the field of multi-domain learning, we thought of using adapter modules to explain the small but significant asymmetry information between OU. In the multi-domain context, the parameters of the larger backbone network are often shared across different domains, while adapter modules with fewer parameters are used to adapt different information patterns between domains \cite{lee2021adaptable,li2022cross}. This aligns closely with the situation we are dealing with.
The specific idea is that the strong correlation between OU implies that the input of OU should share most of the parameters for feature extraction and prediction. However, a small portion of distinct parameters is also needed to extract the asymmetric information. Therefore, the concept of the adapter model can be applied to this situation.

Based on the characteristics of the input images described above, we formulate our network as a bi-channel CNN, where our CNN has a shared backbone for the inputs of the two channels, and distinct adapter modules corresponding to each input channel. 
Currently, in the field of ophthalmic imaging, multi-channel network models commonly take different types or modalities of images of the same eye \cite{ma2021multichannel,yi2023retinal}, which differs from our research setup. 
Moreover, the majority of them generate separate neural networks for each input channel, with fusion of the multiple channels occurring only at the end of fitting the whole network \cite{gour2021multi,xu2022dual}. To our best knowledge, only one article has considered the correlation between fundus image inputs and added a spatial correlation model in their proposed bi-channel network \cite{he2021multi}.

\begin{figure}[tb]
    \centering
\includegraphics[width=0.9\columnwidth]{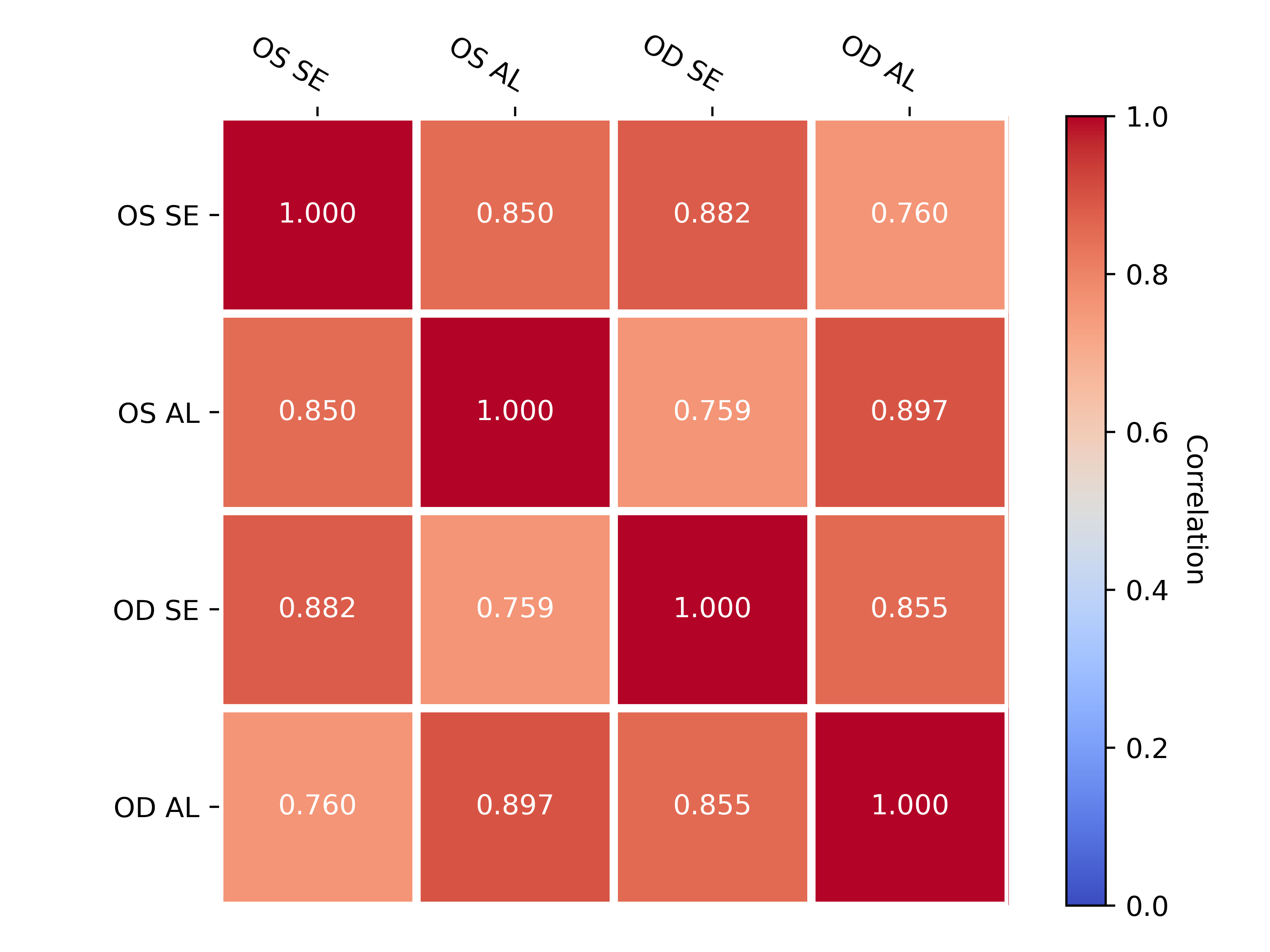}
    \caption{Heatmap of correlation between SE and AL from both eyes}
    \label{fig:heatmap}
\end{figure}

The assessment of myopia has predominantly relied on the Spherical Equivalent (SE), neglecting the potential of Axial Length (AL) as a significant outcome measure \cite{zadnik2015prediction}. Existing literature suggests the predictive prowess of AL in anticipating myopia onset \cite{mutti2007refractive}.
In addition, as Fig.~\ref{fig:heatmap} shows, SE and AL of OS and OD are strongly correlated.
These characteristics motivate us to use the interrelationship and take the correlation information between SE and AL into account to potentially further enhance the predictive capability of our models. To our best knowledge, only one work has utilized both SE and AL to predict myopia status based on fundus images and deep learning models \cite{zhong2023cecnn}. However, this work only focuses on left eyes. In this study, since we have SE and AL for both eyes, we formulate the prediction of four outcomes as a multi-label learning problem. 

Regarding the multi-label learning problem from fundus images, most existing literature focuses on the classification of various retinal diseases \cite{sun2022multi,al2024fundus}. 
If we extend the study of the correlation among labels to other fields, the most extensive one is multi-label classification \cite{wu2023ctranscnn,lai2023single}. 
Although there is abundant research on the correlation between labels in this area, what distinguishes our study is that these investigations primarily focus on the correlation between binary labels in the context of multiple classification tasks while our work is based on modeling the correlation between multiple continuous output variables.

Starting from these preliminary considerations, we propose to treat the prediction of SE and AL of both eyes as a multi-label learning problem (with continous outcomes). We propose a framework of multi-label copula-enhanced CNN Learning with Oculus Uterque (OUCopula) for joint prediction of multiple clinical scores, using UWF fundus images on both eyes. Compared with previous studies, OUCopula can incorporate correlation information between SE and AL, the two outcome variables, into the neural network models. It can simultaneously address interocular asymmetries and the strong correlation between the two eyes, thereby enhancing the predictive capability of our model. Below we highlight our main contributions:

\begin{enumerate}
    
    \item We establish a high-quality annotated dataset containing 5228 OU UWF fundus images. All the images have been desensitized, marked, and graded by experienced doctors. 

    \item We are the first to develop a CNN with joint modeling of OU based on cutting-edge UWF images. The results demonstrate our method’s superiority on the predictive power of myopia screening compared to previous models which only take single-eye inputs.

    \item We construct a new bi-channel multi-label architecture with adapter modules and copula-likelihood loss. Our novel strategy of building separate adapters for OS and OD and sharing the backbone models addresses the issue of interocular asymmetry and high correlation between OU by allowing the majority of parameters to be shared between both eyes while preserving their unique heterogenous characteristics. Also the copula-likelihood loss provides a new way to model the correlation between multiple continuous labels and perform joint regressions.

\end{enumerate}

\section{Related Work}


    


The advent of UWF fundus imaging has revolutionized the field of ophthalmology, offering a broader perspective on retinal pathologies compared to traditional fundus images. Concurrently, the integration of DL techniques has emerged as a powerful tool for the analysis and interpretation of UWF fundus images. Several deep learning approaches have been proposed to predict and detect retinal diseases from either regular fundus images \cite{hemelings2021pathological,dong2022artificial} or UWF fundus images \cite{li2021deep,engelmann2022detecting}. 
Despite the widespread application of DL to analysing fundus images, the majority of research has focused on the diagnosis, classification, and segmentation of diseases. 
Several myopia screening models have been developed but they mainly focus on binary classifications \cite{yang2020automatic,choi2021deep}.
There has been relatively little direct prediction of continuous scores of myopia.

Recently, bi-channel and multi-channel learning have become common in the medical imaging field, due to the increasing availability of medical data from multiple sources and modalities \cite{liu2018joint}.
In the field of ophthalmic imaging, the input channels for most multi-channel models commonly consist of different modalities of images of one eye \cite{ma2021multichannel,yi2023retinal}. 
Most researchers assign distinct CNNs for each input channel, and fuse the multiple channels at the end of fitting the network \cite{gour2021multi,xu2022dual}. 
Furthermore, most current CNN-based fundus imaging studies directly utilize the classic network models developed from other computer vision applications without special optimization of the architecture for medical scenarios.
Regarding the case of OU, we find only one article that has considered the correlation between fundus image inputs and adds a spatial correlation model in their proposed bi-channel network \cite{he2021multi}.
Furthermore, there is a paucity of research considering bilateral asymmetry. To our knowledge, there are only very few studies that use CNN to determine the presence of bilateral asymmetry in fundus images \cite{kang2022asymmetry}. Otherwise, there is no literature that considers bilateral asymmetry in retinal disease prediction. 

Multi-label learning research from fundus images mostly focuses on classifying various retinal diseases \cite{al2024fundus,sun2022multi}, and based on our investigation, no DL research in the field of ophthalmology has considered the correlation between labels.
More generally, for multi-label learning in the computer vision field, most existing literature also considers the multiple classification problem \cite{wu2023ctranscnn,lai2023single,song2018deep,chen2021learning,zhu2020m3la}, and some of them consider the spatial correlation across labels determined by small patches/instances of the image. 
In contrast, OUCopula employs a copula model to incorporate the conditional Pearson correlation across the labels, given the image covariates. 
Based on this copula model, we equip the backbone CNN with a new copula-likelihood loss to enhance the prediction capability.

\section{Methods}

\subsection{Residual adapter modules}\label{subsec:structure_CeCNN}
In this study, we use the residual adapter to adapt to the interocular asymmetry between the two eyes.
The residual adapter was introduced in \cite{rebuffi2018efficient}, and it was initially applied in the field of multi-domain learning. In recent years, it has also been widely used in the Natural Language Processing domain for fine-tuning models and transfer learning. The conventional use of adapters in multi-domain learning involves training different adapters based on the same large pre-trained backbone while freezing all the parameters in the backbone model.

Instead, in this work, we adopted a slightly different architecture to accommodate the presence of high correlation and interocular asymmetries.
In our framework, we want to share most parameters in the network to preserve the common features between OU, while allowing small but distinct parts in the network to contain the heterogeneous information for OS and OD individually. 

\begin{figure*}
\centering
\begin{subfigure}{0.72\textwidth}
    \includegraphics[width=\textwidth]{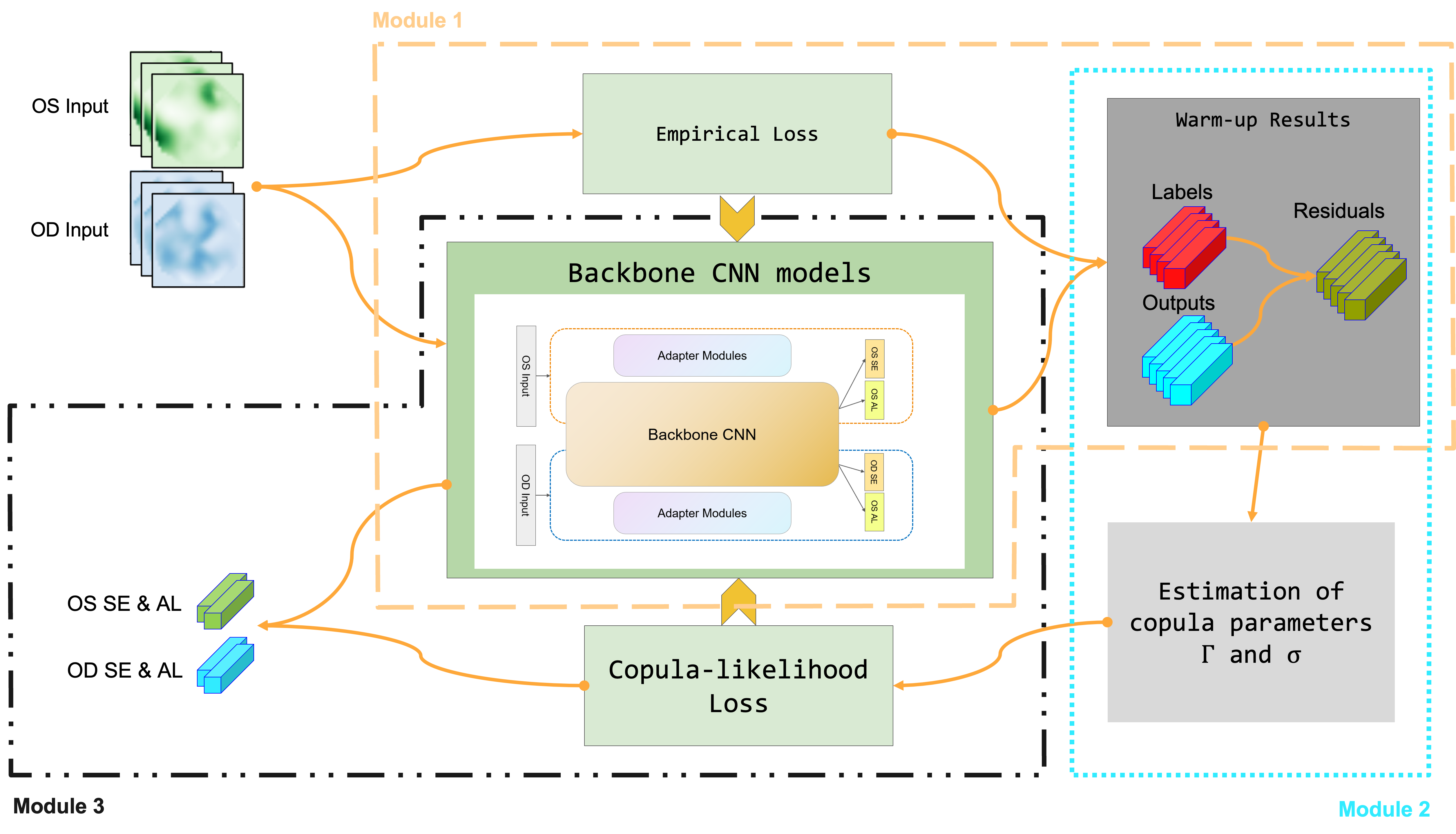}
    \caption{} 
    \label{fig:diagram_OUCopula}
\end{subfigure}
\hfill
\begin{subfigure}{0.263\textwidth}
    \includegraphics[width=\columnwidth]{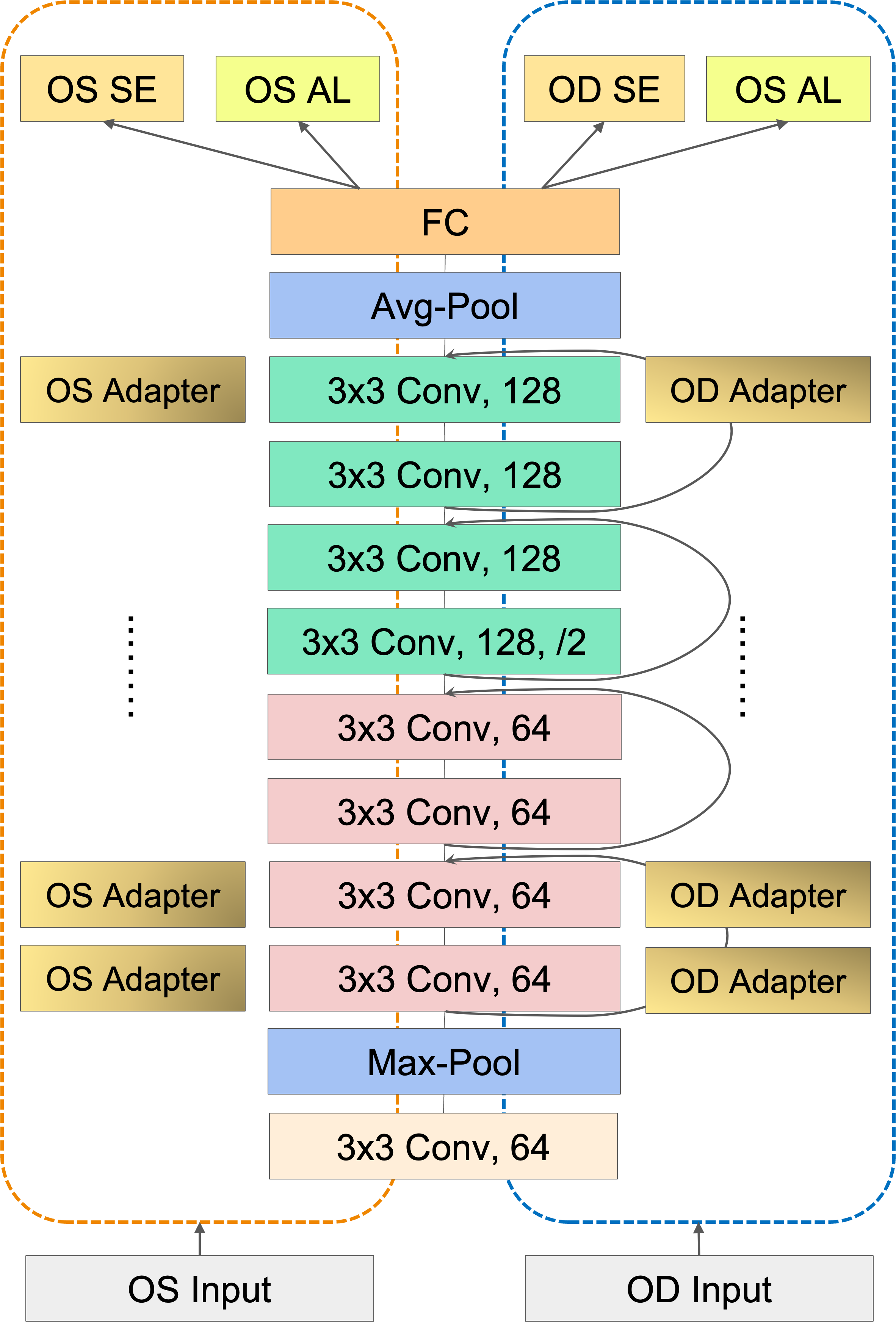}
    \caption{} 
    \label{fig:ResnetArchitect}
\end{subfigure}
\caption{(a) Architecture of proposed OUCopula; (b)Detailed architecture of ResNet backbone with adapter modules.}
\end{figure*}

Hence, we constructed adapter modules for both OS and OD respectively and let them share the backbone model. During training, adapters are updated simultaneously with the backbone model. Specifically, the OS input passes through the backbone model and the OS's adapter, while the OD input goes through the backbone model and the OD's adapter. Each training iteration involves inputting images from both eyes of the same individual. Subsequently, the outputs from both eyes are used together in the computation of the copula-likelihood loss, followed by backward propagation to complete a single training iteration. 
In order to show the structure of the backbone model and adapter modules, and to explain the training process, Fig.~\ref{fig:ResnetArchitect} illustrates the detailed architecture when using ResNet as the backbone.

\subsection{Copula modeling}\label{subsec:copula_modeling}
\subsubsection{Multi-label regression}
Suppose we have $n$ subjects with $J$ different tasks and have at most $K$ labels for each task. 
Let $j =1, \ldots, J$ and $k =1, \ldots, K$ be the task and label indices respectively. 
For the $i$th subject, the covariates are given as $\{X_{ij}\}_{j=1}^J$, where $X_{ij}$ denotes the image corresponding to the $j$th task.
The corresponding labels $\bm{Y}$ are, therefore expressed as $\{y_{ijk}\}, j=1, \ldots, J, k=1, \ldots, K$. 
Under the OU UWF data setting, $J=2$ represents the OS and OD task, $K=2$ represents the AL and SE labels, and $X_{ij}$ denotes the UWF fundus image of either the OS ($j=1$) or the OD $(j=2)$. 
We consider the following multi-label mean regression model 
\begin{align}
\label{multi-res}
y_{ijk} = g_{jk}(X_{ij}) + \epsilon_{ijk}, ~ j=1, 2, ~k=1, 2, 
\end{align}
where $g_{jk}$ denotes the unknown regression function associating the covariate $X_{ij}$ with $y_{ijk}$, and the $\epsilon_{ijk}$ represent noise. 
Without loss of generality, we assume that $E{\epsilon}_{ijk} = 0$ and $\bm{\epsilon}_{i} \perp \bm{\epsilon}_{t}$ for $i \not = t$, where $\bm{\epsilon}_i = (\epsilon_{i11}, \ldots, \epsilon_{iJK})^T$.  

\subsubsection{Gaussian copula}
Let $\bm{\epsilon} = (\epsilon_{11},\ldots, \epsilon_{JK})^T$ be the noise vector and its joint density be $f_{\bm{\epsilon}}$. 
As shown by Fig.~\ref{fig:heatmap}, the labels are indeed dependent. 
Hence, it is unsuitable to assume that $f_{\bm{\epsilon}}$ has a product form. 
To incorporate the dependence structure across the labels, we consider the copula model \cite{sklar1959fonctions}. 
Let $C$ be a $p$-dimensional distribution function (CDF) defined on the rectangle $[0, 1]^p$. 
Any $p$-dimensional  joint distribution function $F$ can be completely specified by its marginal distributions and a copula $C$. 
That is, there exists a copula $C$ such that
$$
F(y_1, \ldots, y_p) = C\{F_1(y_1), \ldots, F_p(y_p)\},
$$
where $F_j$ denotes the marginal CDF of $y_j$ for $j=1, \ldots, p$. 
Since all labels studied in this paper are continuous, we consider the Gaussian copula. 
Let $\Phi$ be the CDF of the standard normal distribution and $\Gamma \equiv (\gamma_{tj})_{p\times p}$ be a correlation matrix, where the elements $\gamma_{tj}$ are defined as
\begin{align}
    \label{correlation}
    \gamma_{tj} = corr(\Phi^{-1}\{F_t(y_t)\}, \Phi^{-1}\{F_j(y_j)\}), 
\end{align}
representing the linear correlation between the two Gaussian scores $(\Phi^{-1}\{F_t(y_t)\}, \Phi^{-1}\{F_j(y_j)\})$.
Equipped with the correlation matrix $\Gamma$, a Gaussian copula is given as 
\begin{align}
\label{GaussCopula}
    C(\bm{y}|\Gamma) = \Phi_p(\Phi^{-1}\{F_1(y_1)\}, \ldots, \Phi^{-1}\{F_p(y_p)\}|\Gamma), 
\end{align}
where $\Phi_p(\cdot|\Gamma)$ denotes the $p$-dimensional standard Gaussian CDF with correlation matrix $\Gamma$. 

\subsubsection{Copula loss}
By taking the derivative of \eqref{GaussCopula}, we obtain the copula density as 
\begin{eqnarray}
\label{copdensity}
    \begin{aligned}
         f_{\epsilon}&(t_{11}, \ldots, t_{JK}) \\
         &\propto \exp\left\{\frac{1}{2} \bm{q}^T(I_{J\times K} -\Gamma^{-1})\bm{q}\right\}
         \prod_{(j, k) = (1, 1)}^{(J, K)} f_{jk}(t_{jk}), 
    \end{aligned}
\end{eqnarray}
where $\bm{q} = \{(\Phi^{-1}(F_{11}(t_{11})), \ldots, \Phi^{-1}(F_{JK}(t_{JK}))\}^T$. 
The copula loss OUCopula model is defined as minus the log-likelihood based on \eqref{copdensity}. 

We often consider the Gaussian case where $\epsilon_{jk} \sim N(0, \sigma_{jk}^2)$. 
In this case, the copula density \eqref{copdensity} reduces to a multivariate Gaussian density, so the copula loss is given by 
\begin{eqnarray}
    \label{GaussianLoss}
    \begin{aligned}
        \mathcal{L}(g_{11}, &\ldots, g_{JK}|\bm{Y}) = \\ 
&-\sum_{i=1}^n \log N_{J \times K}(e_{i11}, 
\ldots, 
e_{iJK}; \bm{0}_{J\times K}, \Sigma), 
    \end{aligned}
\end{eqnarray}
where $e_{ijk} = y_{ijk} - g_{jk}(X_{i})$ represent the residuals, and $\Sigma \equiv (\Sigma)_{(jk) = 11}^{(jk)=JK} = \sigma_{j_1k_1} \sigma_{j_2k_2} \gamma_{j_1k_1, j_2k_2}$ represents the covariance matrix.

\subsection{Choice of backbone models}\label{subsec:choice_backbone}
The copula loss is flexible and generalizable in the sense that it does not impose specific requirements on the choice of the backbone model. We opted to use ResNet, a classic, concise, and versatile structure, as the backbone model.
As indicated in \cite{zhong2023cecnn}, applying copula-likelihood loss should avoid overfitting during training; otherwise, it would fail to extract effective dependence information from overfitted residuals. Due to the relatively small size of our UWF dataset  (a few thousand images), we followed the approach applied in  \cite{zhong2023cecnn}: using a simplified ResNet18 (last two CNN blocks removed) as the backbone model.

\subsection{End-to-end OUCopula}
In this section, we describe our OUCopula framework, which consists of three modules: (1) a warm-up module with adapter modules that trains the backbone CNN under empirical losses; (2) a copula estimation module that estimates the parameters in the Gaussian copula based on the results of the
warm-up module; (3) the OUCopula module that trains the backbone CNN using the derived copula-likelihood loss that can efficiently capture the dependence structure across multiple labels through a Gaussian copula model. We also summarize the OUCopula in Algorithm \ref{alg:OUCopulareg} and show an overview of the OUCopula architecture in Fig.~\ref{fig:diagram_OUCopula}.

In detail, Module 1 includes the basic construction of backbone CNNs under empirical losses. Residual adapters are applied in the backbone CNN as described in \ref{subsec:structure_CeCNN}.
Then all the numerous parameters included in the CNN layers are updated by the Adam algorithm \cite{kingma2014adam} to optimize the losses. 
In Module 2, we  estimate the two copula parameters including the marginal sample standard deviation (SDs) $\sigma_{jk}$  and the Pearson correlation $\gamma_{j_1k_1, j_2k_2}$ by the sample SD and sample Pearson correlation respectively.
Finally, in Module 3, we determine the copula-likelihood loss based on the estimated copula parameters and use the backbone CNNs from Module 1 to train the OUCopula under the copula-likelihood loss. 
Note that in this Module, OUCopula does not rely on any specific architecture in the backbone CNN, and thus, there is great flexibility that we can use any backbone CNNs.

\begin{algorithm}[tb]
\caption{End-to-end OUCopula}\label{alg:OUCopulareg}
\begin{algorithmic}[1]
\Require Training data $\{\mathcal{X}_i, \bm{Y}_i = (y_{i 11}, \ldots, y_{iJK})\}_{i=1}^n$. 
\Ensure Trained OUCopula $(\hat{g}_{11}, \ldots, \hat{g}_{JK})$. 
\Algphase{Module 1: Warm-up CNN}
\For{$j = 1, \ldots, J, k=1, \ldots, K$}
\State $g_{jk} =$ \text{Backbone CNN with adapter modules}. 
\State $\hat{g}_{jk}^0 = \arg \min n^{-1} \sum_{i=1}^n (y_{ijk} - g_{jk}(\mathcal{X}_i))^2$. 
\EndFor
\Algphase{Module 2: Copula parameters estimation}
\State Obtain residuals $e_{ijk} = y_{ijk} - \hat{g}_{jk}^0(\mathcal{X}_i) $. 
\State $\hat{\sigma}_{jk} = \text{sd}(\bm{e}_{ijk})$, $\hat{\gamma}_{j_1k_1, j_2k_2} = corr(\bm{e}_{ij_1k_1}, \bm{e}_{ij_2k_2})$. 
\Algphase{Module 3: OUCopula}
\State $g_{jk} =$ \text{Trained backbone CNN in Module 1}.
\State Define the loss $\mathcal{L}_2$ the same as \eqref{GaussianLoss}. 
\State $(\hat{g}_{11}, \ldots, \hat{g}_{JK}) = \arg \min \mathcal{L}_2$. 
\end{algorithmic}
\end{algorithm}

\section{Experiments}
In this section we evaluate our method quantitatively against the  simplified ResNet backbone model, and investigate whether the proposed framework can improve the backbone model by extracting correlation information for the outcomes and addressing the issue of interocular asymmetry.
We will test and compare the results between the sole ResNet model, ResNet + Adapters, and ResNet + Adapters + copula model (OUCopula).

\subsection{Dataset}
The data collection process involved capturing 5228 UWF fundus images from the eyes of 2614 patients using the Optomap Daytona scanning laser ophthalmoscope (Daytona, Optos, UK). All enrolled patients sought refractive surgery treatment and were exclusively myopia patients. The data collection period extended from December 2014 to June 2020, and was conducted at The Eye and ENT Hospital of Fudan University. For this study,  only central gazed images were included and blurred or eccentric images were excluded. The UWF fundus images obtained during the study were exported in JPEG format and compressed to a resolution of 224 x 224 pixels to facilitate subsequent analysis.

\subsection{Experiment Setup}
The dataset of 5228 fundus images was partitioned into the training data set, the validation set, and the testing set, with a ratio of 6:2:2. To mitigate bias in model evaluation and obtain more reliable estimates of the results, 5-fold cross-validation (CV) was employed. Since the outcome variables are all continuous, we calculated the Mean Square Error (MSE) for the results. 
In addition to the total MSE loss of OU, we also performed various sub-analyses to study the detailed pattern of predictive performance. Specifically, the following MSE losses were compared: 

\begin{enumerate}
    
    \item MSE loss for SE and AL of each eye (OS SE, OS AL, OD SE, and OD AL);

    \item Total loss for each eye (OS SE+AL, OD SE+AL); 

    \item Total loss for SE and AL separately (OU SE, OU AL);

    \item Total loss for OU (OU SE+AL).

\end{enumerate}

We compared the predictive performance between the backbone model, the backbone model with adapter modules, and the backbone model with both the adapter modules and copula-likelihood loss (OUCopula), aiming to show that the proposed framework could enhance the performance of the backbone models. The simplified ResNet model mentioned in subsection \ref{subsec:choice_backbone} served as the backbone model for our experiment.
For the single backbone model of ResNet, since there is no adapter module, the bi-channel input was simplified to a single-channel input. This is equivalent to treating the left and right eyes independently as different individual input samples, which is the common approach of using single eye inputs in the majority of the literature dealing with fundus images.
Therefore, comparing the backbone + adapters with the single backbone model can be considered as a comparison between the joint modeling of OU and the traditional modeling of single eyes.

\subsection{Results}\label{subsec:results}
\begin{figure*}[!htb]
    \centering\includegraphics[width=.8\textwidth]{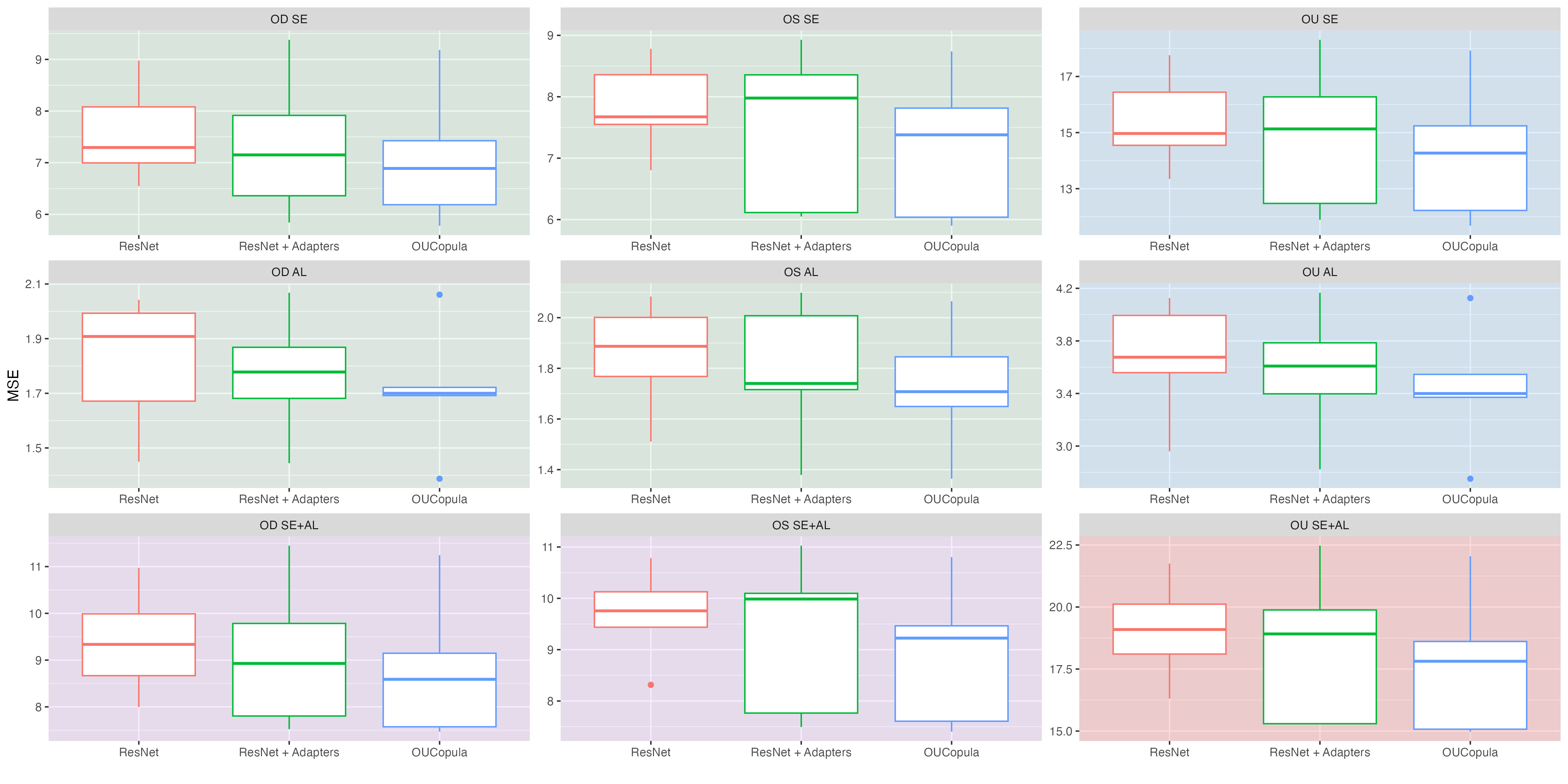}
    \caption{Box plots of 5-fold CV prediction performances of OUCopula and baselines for all sub-results. 
    } \label{fig:result_resnet_boxplots}
\end{figure*}

\begin{figure*}[!htb]
    \centering
    \includegraphics[width=0.8\textwidth]{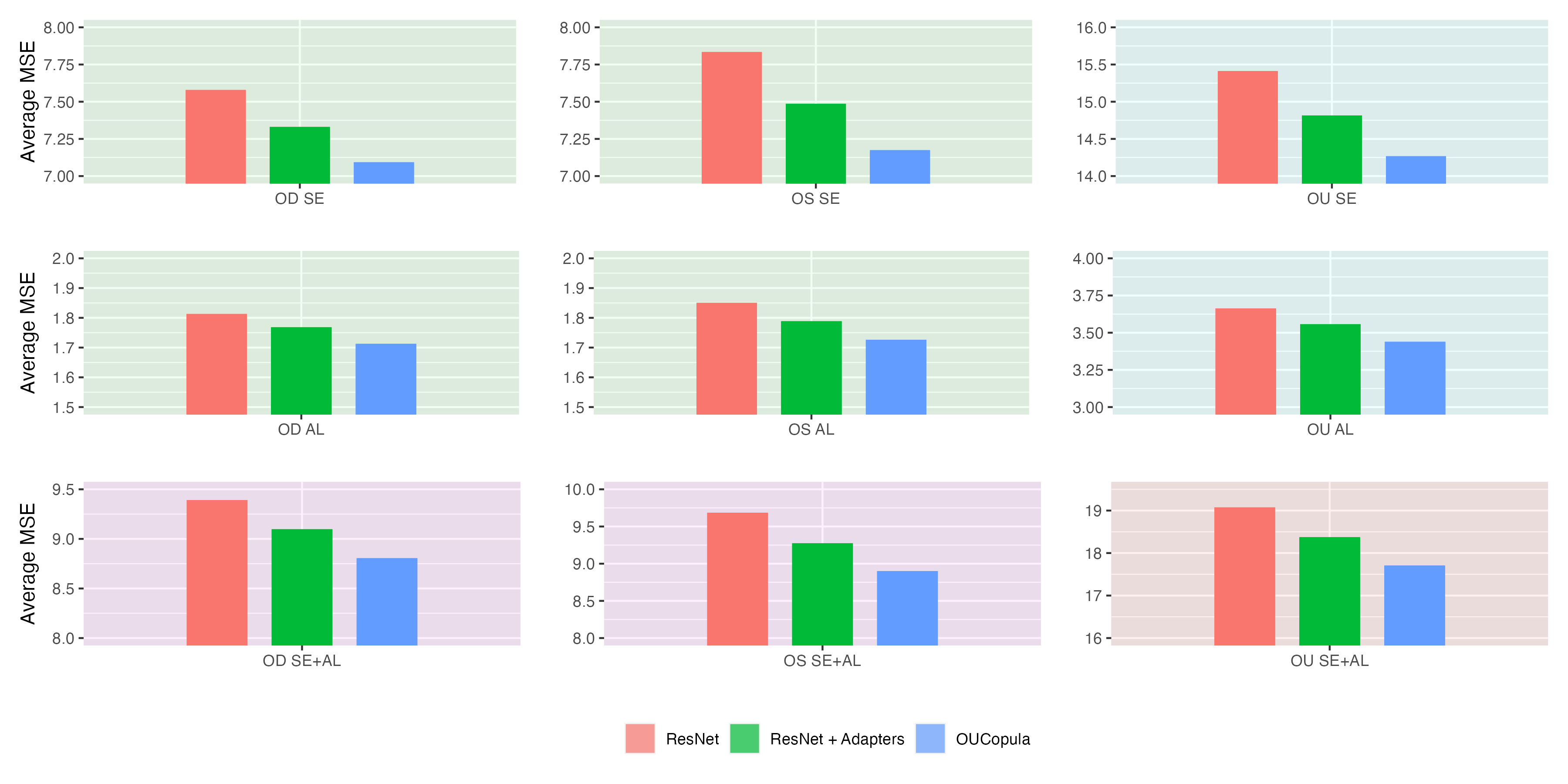}
    \caption{The prediction performances (in average among 5 folds) of OUCopula and baselines for all sub-results.} 
    \label{fig:result_resnet_histogram}
\end{figure*}

Fig.~\ref{fig:result_resnet_boxplots} and Fig.~\ref{fig:result_resnet_histogram} compare OUCopula and baseline models.
In these figures, the charts with a green background represent the results for a single label in one eye, those with a light blue background represent the total results for a single label, those with a light purple background represent the total results for one eye, and those with a red background represent the overall results for both eyes.
In these two 3 $\times$ 3 charts, each third chart in every row or column is the sum of the preceding two charts.

According to Fig.~\ref{fig:result_resnet_boxplots} and Fig.~\ref{fig:result_resnet_histogram}, on average OUCopula surpasses both pure ResNet and ResNet + Adapters on almost all sub-results. Whether it's the MSE loss for the SE or AL of a single eye, or the total loss for both the left and right eyes, OUCopula achieves the best performance. We also have the following observations:  

\begin{enumerate}
\item If we compare ResNet alone with ResNet + Adapters, we see an average improvement in total loss of 3.69\%. Comparing ResNet alone with OUCopula, the total loss shows an average improvement of 7.18\%. It can be roughly observed that the contributions of the residual adapters and copula loss to the prediction improvement are of a similar level, with adapter modules contributing slightly more. 

\item Among all five folds, ResNet + Adapters shows improvements in 3 out of 5 folds, but not in every fold; on the other hand, OUCopula performed better than ResNet in 4 out of 5 folds. Additionally, OUCopula consistently outperforms ResNet + Adapters in most sub-results.

\item Whether we compare the results for SE or AL separately, or compare the overall results for OS, OD or OU, OUCopula shows a relatively balanced improvement of around 6\%-8\% on average. Specifically, utilizing OUCopula improves the MSE by 8.42\%	for OS SE, 6.69\% for OS AL, 6.41\% for OD SE, 5.54\% for OD AL, 8.09\% for OS SE+AL, 6.25\% for OD SE+AL, and 7.18\% for OU SE+AL. The improvements on SE are slightly better than those on AL. This phenomenon differs from previous literature where copula-likelihood loss in \cite{zhong2023cecnn} tended to optimize outcomes with smaller variances, and the RMSE of SE was improved by 2.232\% and the RMSE of AL was improved by 3.683\% on average using the same ResNet backbone.

\end{enumerate}

\section{Discussion}

Overall, the experimental results show that OUCopula consistently outperforms both the backbone model and backbone + Adapter across various sub-results, indicating that OUCopula exhibits remarkable efficacy in enhancing the predictive accuracy of backbone models. 
Observations presented in subsection \ref{subsec:results} also provide valuable insights into the effectiveness of OUCopula: 
 

\begin{itemize}

\item The improvement of ResNet + Adapters over single backbone models shows that our joint modeling under the bi-channel architecture with residual adapter modules does capture the interocular asymmetries (the heterogeneity of OU inputs), which was often ignored in the literature.
Meanwhile, OUCopula consistently surpasses ResNet + Adapters across all sub-results among most folds, indicating that our copula modeling correctly characterizes the conditional correlation between the labels (given the image covariates).

\item Compared to the backbone + Adapters, OUCopula shows improvement across all sub-results among most folds, indicating the robustness and the stable performance of OUCopula in cross validation. 

\item An interesting finding is that OUCopula can improve the prediction of multiple labels in a balanced way regardless of its marginal SD; say, the improvements on SE tend to be slightly more pronounced than those on AL, while the marginal SD of SE is significantly larger than that of AL.
This phenomenon is different from the common results in the modeling of task-dependent uncertainty (e.q. (7) of \cite{kendall2018multi} for instance). 
We conjecture that, our more complex framework of jointly modeling OU and multiple labels allows our model to extract a broader range of information from input images and mitigate the square loss caused by larger marginal SD, leading to more balanced multi-label predictions.


\end{itemize}

Last but not least, we want to discuss the adaptability of OUCopula.
While the current focus is on predicting four outcome labels, our model's architecture is not constrained by this specific number. From subsection \ref{subsec:copula_modeling} we see that the inherent flexibility of copula-likelihood loss allows for straightforward extensions to handle a more extensive range of outcome labels. This scalability is particularly relevant in medical applications where comprehensive assessments involving multiple clinical parameters are common. Researchers can extend the model to handle more than four outcomes for more comprehensive prediction and complex clinical practices. 
Furthermore, it is essential to highlight that our bi-channel architecture is designed for flexibility and can readily be extended to a multi-channel model. This adaptability is crucial, especially in scenarios where additional imaging modalities or channels may provide complementary information. Future work could explore the integration of diverse data sources to further enhance the model's predictive capacity.
Additionally, since the dataset is relatively small (only a few thousand input images), we adopted the simple solution for overfitting (simplifying the backbone network) mentioned in \cite{zhong2023cecnn}. The characteristics of our copula-likelihood loss mean that we can apply various kinds of backbone models in the OUCopula framework. Hence, exploring how to apply OUCopula to more advanced and much larger models while avoiding overfitting can be considered valuable directions for future work.

\section{Conclusion}
Our study has demonstrated the effectiveness of the OUCopula model in leveraging UWF eye images for myopia screening. The bi-channel input structure and the integration of adapter modules allowed for simultaneous modeling of OU, addressing both the complex interocular asymmetry and correlation between them. The ability of the copula-likelihood loss to incorporate the dependence structure between multiple continuous outcome labels is a notable strength. Furthermore, the adaptability of OUCopula means that it can be generalized to different kinds of backbone models, and extended to multi-channel, multi-label problems.





\bibliographystyle{named}
\bibliography{main}

\end{document}